\documentclass{appolb}
\usepackage{epsfig}

\begin{document}
\title{Structure of Scalar Mesons $f_{0}(600)$, $a_{0}(980)$, $f_{0}(1370)$ and $a_{0}(1450)$%
\thanks{Presented by D. Parganlija at the Excited QCD Workshop, 31.1.-6.2.2010, in Tatranska Lomnica (Slovakia)}}%
\author{Denis Parganlija$^{a}$, Francesco Giacosa$^{a}$ and Dirk H. Rischke$^{a,b}$
\address{$^a$Institute for Theoretical Physics, Goethe University,
Max-von-Laue-Str.\ 1, D--60438 Frankfurt am Main, Germany \and $^b$Frankfurt Institute for Advanced Studies, Goethe University,
Ruth-Moufang-Str.\ 1, D--60438 Frankfurt am Main, Germany} }
\maketitle
\begin{abstract}
We calculate low-energy meson decay processes and pion-pion scattering lengths in a globally chirally invariant two-flavour linear sigma model exploring the quark content of the scalar mesons $f_{0}(600)$ and $a_{0}(980)$, as well as $f_{0}(1370)$ and $a_{0}(1450)$. To this end, we investigate which one of these sets of fields is more likely to contain quark-antiquark states.
\end{abstract}
\PACS{12.39.Fe, 13.20.Jf, 13.25.Jx, 13.75.Lb}
  
\section{Introduction}
Mesons and baryons at low energies can be described using various effective field theories and models that contain hadrons as degrees of freedom rather than quarks and gluons. These effective approaches may be based on the linear \cite{gellmanlevy} or the non-linear \cite{weinberg} realisation of the $U(N_{f})_{R}\times U(N_{f})_{L}$ (the so-called \textit {chiral}) symmetry of the Quantum Chromodynamics (QCD) with $N_f$ flavours. The chiral symmetry is broken in two ways: spontaneously, by the presence of the quark condensate $\langle\bar{q}q\rangle=\langle\bar{q}_{R}q_{L}+\bar{q}_{L} q_{R}\rangle\neq 0$ \cite{SSB}, and explicitely, by the non-zero, non-degenerate quark masses as well as the axial anomaly \cite{Hooft}. In this paper, we present a linear sigma model with vector and axial-vectors mesons and global chiral symmetry \cite{Reference1,Mainz,Paper1}. 

Our model contains two scalar fields, $\sigma$ and $a_0$ $[\sigma =\frac{1}{\sqrt{2}}(\bar{u}u+\bar{d}d)$, $%
a_{0}^{0}=\frac{1}{\sqrt{2}}(\bar{u}u-\bar{d}d)]$. As outlined in Refs.\ \cite{Mainz,Paper1}, the fields $\sigma$ and $a_0$ can be assigned to physical fields according to two different scenarios: (\textit{i}) they can be identified with the states $f_{0}(600)$ and $a_{0}(980)$ [that form a part of a larger $\bar{q}q$ nonet which comprises $f_{0}(980)$, $a_{0}(980)$, $\kappa(800)$ and $f_{0}(600)$ -- resonances below 1 GeV]; (\textit{ii}) they can be identified with the states $f_{0}(1370)$ and $a_{0}(1450)$ forming a part of a $\bar{q}q$ nonet (plus a glueball) that comprises $f_{0}(1370)$, $f_{0}(1500)$, $f_{0}(1710)$, $a_{0}(1450)$, $K_{0}(1430)$ -- resonances above 1 GeV (see also Ref.\ \cite{refs1}). Scalar mesons below 1 GeV, whose spectroscopic wave functions possibly contain a dominant tetraquark contribution \cite{refs2}, may be introduced in the second scenario as additional scalar fields or may arise as mesonic-molecular states because of interactions of pseudoscalar mesons \cite{Isgur}.

In this paper, we will describe briefly the consequences of the assignments (\textit{i}) and (\textit{ii}). The paper is organised as follows: in Sec.\ \ref{2} the model is introduced,
in Sec.\ \ref{3} the results for both scenarios are presented and
in Sec.\ \ref{4} we present our conclusions. More detailed
results are described in Ref.\ \cite{Paper1}.

\section{The Model} \label{2}
The Lagrangian of the globally invariant linear sigma model with
$U(2)_{R} \times U(2)_{L}$ symmetry reads
\cite{Reference1,Mainz,Paper1,Boguta}:
\begin{eqnarray}
\lefteqn{\mathcal{L}=\mathrm{Tr}[(D^{\mu}\Phi)^{\dagger}(D^{\mu}\Phi)]-m_{0}^{2}
\mathrm{Tr}(\Phi^{\dagger}\Phi)-\lambda_{1}[\mathrm{Tr}(\Phi^{\dagger}%
\Phi)]^{2} -\lambda_{2}\mathrm{Tr}(\Phi^{\dagger}\Phi)^{2}}\nonumber\\
&&  -\frac{1}{4}\mathrm{Tr}[(L^{\mu\nu})^{2}+(R^{\mu\nu})^{2}]+\frac{m_{1}^{2}%
}{2} \mathrm{Tr}[(L^{\mu})^{2}+(R^{\mu})^{2}]+\mathrm{Tr}[H(\Phi+\Phi
^{\dagger})]\nonumber\\
&&  +c(\det\Phi+\det\Phi^{\dagger})-2ig_{2}(\mathrm{Tr}\{L_{\mu\nu}[L^{\mu
},L^{\nu}]\} +\mathrm{Tr}\{R_{\mu\nu}[R^{\mu},R^{\nu}]\})\nonumber\\
&&  -2g_{3}\left[  \mathrm{Tr}\left(  \left\{  \partial_{\mu}L_{\nu}-ieA_{\mu
}[t^{3},L_{\nu}] +\partial_{\nu}L_{\mu}-ieA_{\nu}[t^{3},L_{\mu}]\right\}
\{L^{\mu},L^{\nu}\}\right)  \right. \nonumber\\
&& +\left.  \mathrm{Tr} \left(  \left\{  \partial_{\mu}%
R_{\nu}-ieA_{\mu}[t^{3},R_{\nu}] +\partial_{\nu}R_{\mu}-ieA_{\nu}[t^{3}%
,R_{\mu}]\right\}  \{R^{\mu},R^{\nu}\}\right)  \right] \nonumber\\
&&  +\frac{h_{1}}{2}\mathrm{Tr}(\Phi^{\dagger}\Phi)\mathrm{Tr}[(L^{\mu})^{2}
+(R^{\mu})^{2}]+h_{2}\mathrm{Tr}[(\Phi R^{\mu})^{2}+(L^{\mu}\Phi)^{2}] \nonumber \\
&& + 2h_{3}\mathrm{Tr}(\Phi R_{\mu}\Phi^{\dagger}L^{\mu})+... \label{Lagrangian}%
\end{eqnarray}
where $\Phi=(\sigma+i\eta_{N})\,t^{0}+(\vec{a}_{0}+i\vec{\pi})\cdot\vec{t}$
(scalar and pseudoscalar degrees of freedom; our model is
currently constructed for $N_{f}=2$ - thus, our eta meson $\eta_{N}$ contains
only non-strange degrees of freedom), $L^{\mu}=(\omega^{\mu}%
+f_{1}^{\mu})\,t^{0}+(\vec{\rho}^{\mu}+\vec{a}_{1}^{\mu})\cdot\vec{t}$ and
$R^{\mu}=(\omega^{\mu} - f_{1}^{\mu})\,t^{0}+(\vec{\rho}^{\mu} - \vec{a}_{1}^{\mu
})\cdot\vec{t}$ (vector and axial-vector degrees of freedom); $t^{0}$, $\vec{t}$ are the generators of
$U(2)$; $D^{\mu}\Phi=\partial^{\mu}\Phi-ig_{1}(L^{\mu} \Phi -\Phi R^{\mu})-ieA^{\mu}[t^{3},\Phi]$ ($A^{\mu}$ is the photon field), $L^{\mu\nu
}=\partial^{\mu}L^{\nu}-ieA^{\mu}[t^{3},L^{\nu}]-(\partial^{\nu}L^{\mu
}-ieA^{\nu}[t^{3},L^{\mu}])$, $R^{\mu\nu}=\partial^{\mu}R^{\nu}-ieA^{\mu
}[t^{3},R^{\nu}]-(\partial^{\nu}R^{\mu}-ieA^{\nu}[t^{3},R^{\mu}])$. The dots
refer to further globally invariant terms of naive scaling dimension four that are not important in the following.
The explicit breaking of the global symmetry is described by the term
Tr$[H(\Phi+\Phi^{\dagger})]\equiv h\sigma$ $(h=const.)$ and the chiral anomaly by the term $c\,(\det\Phi+\det\Phi^{\dagger})$ \cite{Hooft}. The reason to restrict the consideration to operators up to fourth order may be found in Ref.\ \cite{dynrec}.

The Lagrangian (\ref{Lagrangian}) contains 12 parameters: $\lambda_{1}$, $\lambda_{2}$,
$c$, $h_{0}$, $h_{1}$, $h_{2}$, $h_{3}$, $m_{0}^{2}$, $g_{1}$, $g_{2}$, $g_{3}$,
$m_{1}$. The parameter $g_3$ is not important for the following. Six parameters can be determined using the masses $m_{\pi}$, $m_{\eta_{N}}$ (about $700$ MeV, obtained by 'unmixing' the physical $\eta$ and $\eta^{\prime}$ mesons), $m_{\rho}$ and $m_{a_{1}}$, the pion decay constant $f_{\pi}$ (via the Eq. $\phi=Zf_{\pi}$) and the experimentally well-known decay width $\Gamma_{\rho\rightarrow\pi\pi}= (149.1\pm0.8)$ MeV \cite{PDG}. One more parameter can be determined from the scenario-dependent value of $m_{a_{0}}$ set to $m_{a_0(980)}=980$ MeV in Scenario I and to $m_{a_0(1450)}=1474$ MeV in Scenario II. All quantities of interest can be then expressed using four parameters, which for convenience are chosen to be $m_{\sigma}$, $Z$, $m_{1}$ and $h_2$.

\section{Results and discussions} \label{3}
\subsection{Scenario I: Scalar Quarkonia below 1 GeV}
\begin{figure}[t]
\centering
\includegraphics[scale = 0.32]{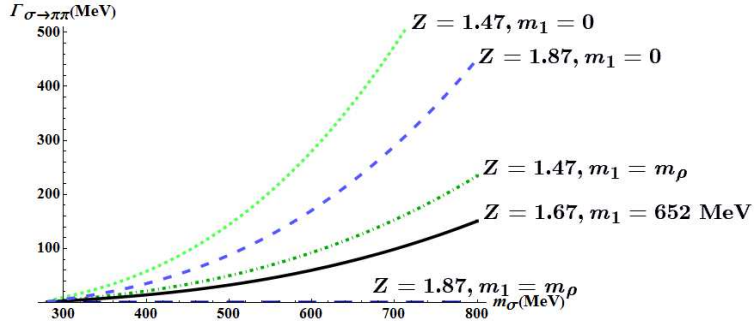}
\caption{$\Gamma_{\sigma \rightarrow \pi \pi}$ as function of $m_\sigma$. The PDG \cite{PDG} quotes $\Gamma_{\sigma
\rightarrow \pi\pi}=(600-1000)$ MeV.}
\label{Figure1}
\end{figure}

The number of independent parameters can be decreased using the decay width 
$\Gamma_{f_{1}\rightarrow a_0(980)\pi}= (8.748 \pm 2.097)$ MeV which in our model depends on $Z$ and $h_2$ and thus allows us to express $h_2$ via $Z$ (the results are virtually independent of the experimental uncertainty of the decay width). We obtain $h_2 \sim 80$. Consequently, we are left with three independent parameters $Z$, $m_{1}$ and $m_{\sigma} \equiv m_{f_0(600)}$. These parameters can be calculated using the $\chi^2$ method \cite{Paper1} considering the decay width $a_{1}\rightarrow\pi\gamma$ [the experimental value is $\Gamma_{a_1 \rightarrow \pi \gamma} = (0.640 \pm 0.246)$ MeV \cite{PDG}] and the $\pi\pi$ scattering lengths $a_{0}^{0}$ and $a_{0}^{2}$ [with the values $a_{0}^{0}=0.218\pm0.020$ and $a_{0}^{2}=-0.0457\pm0.0125$ \cite{Peyaud}]. \\
In this way we obtain $Z=1.67\pm0.2$. The parameter $m_1$ has large uncertainties in this scenario. These can be further constrained using the following consideration: $m_1$ is a part of the $\rho$ mass term
\begin{equation}
m_{\rho}^{2}=m_{1}^{2}+\frac{\phi^{2}}{2}[h_{1}+h_{2}(Z)+h_{3}(Z)]
\label{rhomass}
\end{equation}
and can be related to the gluon condensate. The term $\phi^{2}[h_{1}+h_{2}(Z)+h_{3}(Z)]/2$ in Eq.\ (\ref{rhomass}) represents the contribution of the chiral condensate. Note that $h_{2}\equiv h_{2}(Z)$ and $h_{3}\equiv h_{3}(Z)$ are functions of $Z$ \cite{Paper1}. We consider $m^2_1$ as varying between 0 and $m_\rho^2$: in fact, a negative $m^2_1$ would imply that the vacuum is not stable in the limit $\phi\rightarrow 0$; $m_{1}^{2}>m_{\rho}^{2}$ would imply that the contribution of the chiral condensate is negative. This is counter-intuitive and at odds with various microscopic approaches such as the NJL model. Combining the condition $m_1 \in [0,m_\rho]$ with the results from our $\chi^2$ calculation then yields $m_1 = (652_{-652}^{+123})$ MeV. Finally, the scattering length $a_0^0$ leads to $m_{\sigma}=332_{-44}^{+145}$ MeV \cite{Paper1}.\\
With these values of $Z$, $m_1$ and $m_\sigma$, we obtain the decay width $\Gamma_{\sigma \rightarrow \pi \pi}$, see Fig.\ \ref{Figure1}. The values of $\Gamma_{\sigma \rightarrow \pi \pi}$ are too small when compared to the PDG
data \cite{PDG}. Even for the largest value $m_\sigma = 477$ MeV allowed by the scattering length $a_0^0$ we obtain at most $\Gamma_{\sigma\rightarrow\pi\pi} \cong 145$ MeV. In all
other cases, the decay width is smaller. Hence, we conclude that the isoscalar meson in our model is most likely not $f_{0}(600)$, thus disfavouring this resonance [and consequently also $a_0(980)$] as a predominantly $\bar{q}q$ state.

\subsection{Scenario II: Scalar Quarkonia above 1 GeV}
\begin{figure}[t]
\centering
\includegraphics[scale = 0.31]{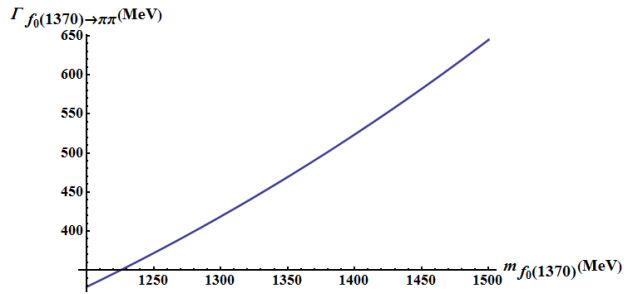}
\caption{$\Gamma_{f_0(1370) \rightarrow \pi \pi}$ as function of $m_{f_0(1370)}$. The PDG \cite{PDG} quotes the mass in the range 1200--1500 MeV and the width between 200 and 500 MeV.}
\label{f0pionpionf}
\end{figure}

The aforementioned problem of the unphysically small
two-pion decay width of the sigma meson may be resolved by identifying the fields 
$\sigma$ and $a_{0}$ of our model with the resonances $f_{0}(1370)$ and $a_{0}(1450)$, respectively.\\
The number of free parameters ($m_{\sigma}$, $Z$, $m_{1}$ and $h_2$) can be decreased expressing the parameter $h_2$ via $Z$ using the total decay width $\Gamma_{a_0 \equiv a_0(1450)} = (265 \pm 13)$ MeV \cite{PDG} (we obtain $h_{2} \simeq 5^{+5}_{-30}$). Combining the decay width $\Gamma_{a_1 \rightarrow \pi \gamma} = (0.640 \pm 0.246)$ MeV \cite{PDG}, that in our model depends only on $Z$, with the condition $m_1 \leq m_\rho$ in Eq.\ (\ref{rhomass}), where the large-$N_c$ suppressed $h_1 \equiv 0$ \cite{Paper1}, we obtain $Z=1.67_{-0.07}^{+0.21}$ and $m_{1}=720_{-140}^{+55}$ MeV. Therefore, the parameter $m_1$ is much better constrained in Scenario II than in Scenario I; the gluon condensate appears to be dominant in $m_\rho$. We are thus left with one free parameter, $m_\sigma \equiv m_{f_0(1370)}$, that will be varied within the experimentally known band \cite{PDG} in order to ascertain whether our result for $\Gamma_{f_{0}(1370)\rightarrow\pi\pi}$ is in agreement with experimental data.\\
The two-pion decay width of $f_0(1370)$ is represented in Fig.\ \ref{f0pionpionf}. Under the assumption that $f_0(1370)$ predominantly decays into two pions \cite{buggf0}, we obtain a good agreement with the experimental values \cite{PDG} if $m_{f_{0}(1370)} \leq 1380$ MeV. Other contributions to the decay
width may reduce this upper bound on $m_{f_0(1370)}$ but nonetheless the correspondence with the experimental data is apparently a lot better in this scenario. Therefore, our results favour $f_0(1370)$ as a (predominantly) scalar quarkonium. Then the $a_{0}(1450)$ meson (the decay width of which was used to express $h_{2}$ via $Z$) is also favoured to be predominantly a $\bar{q}q$ state. Note that the results for the four-pion decay channel of $f_0(1370)$ which follow from the Lagrangian (\ref{Lagrangian}) are also in a good agreement with the experimental data (see Ref.\ \cite{Paper1}).

\section{Conclusions} \label{4}
A linear sigma model with vector and axial-vector mesons and 
global chiral invariance has been presented. The assignment of the fields in the model to the physical states is straightforward in all cases with the exception of the scalars $\sigma$ and $a_0$. They can be assigned either to $f_{0}(600)$ and $a_{0}(980)$ (Scenario I) or to $f_{0}(1370)$ and $a_{0}(1450)$ (Scenario II).\\
As shown in the previous section, Scenario I fails to describe the two-pion decay width of the $f_0(600)$ meson thus disfavouring it, and also $a_0(980)$, as (predominantly) $\bar{q}q$ states. The decay modes of other particles that can be considered in this scenario are, however, in agreement with experimental data \cite{Paper1}.\\
If the scalar fields $\sigma$ and $a_{0}$ are assigned to the
$f_{0}(1370)$ and $a_{0}(1450)$ states, respectively, then the result for the two-pion decay width of the isoscalar meson is considerably improved. We obtain
$\Gamma_{f_{0}(1370)\rightarrow\pi\pi}\simeq (300$-$500)$ MeV for
$m_{f_{0}(1370)}= (1200$-$1400)$ MeV (see Fig.\ \ref{f0pionpionf}). Thus, the assertion that the scalar mesons above 1 GeV, $f_{0}(1370)$ and $a_{0}(1450)$, are (predominantly) $\bar{q}q$ states appears to be favoured by the experimental data.\\
As outlined in Ref.\ \cite{Paper1}, future work should include the mixing of the quarkonia with a glueball as well as calculations in the $N_f=3$ sector \cite{PGR} and at nonzero temperatures and chemical potentials (such as those of Ref.\ \cite{RS}) in order to study the chiral transition.

\end{document}